\documentclass[12pt]{article}

\usepackage{putex}
\usepackage{graphicx}
\usepackage{latexsym,amsmath,amsfonts,amssymb}
\usepackage{bbm}
\usepackage{hyperref}

\def\btab{\begin{table}[h] \begin{center} \begin{tabular}{l lp{3in}}}
      \def\etab{\end{tabular} \end{center} \end{table}}
\def\btabm{\begin{center} \begin{tabular}}
    \def\etabm{\end{tabular} \end{center}}
\def\eg{{\it e.g.}}

\def\ie{{\it i.e.}}

\def\m{{\mu}}
\def\n{{\nu}}

\def\s{\sqrt}

\def\we{\wedge}

\def\CN{{\cal N}}
\def\CO{{\cal O}}

\begin{document}

\title{The Chiral Heat Effect}

\authors{Taro Kimura$^{\diamondsuit,\clubsuit}$, and Tatsuma Nishioka$^\spadesuit$}

\institution{UT}{
${}^\diamondsuit$Department of Basic Science, University of Tokyo,
Tokyo 153-8902, Japan
}
\institution{RI}{
${}^\clubsuit$Mathematical Physics Lab., RIKEN Nishina Center, 
Saitama 351-0198, Japan
}
\institution{PU}{
${}^\spadesuit$Department of Physics, Princeton University, Princeton, NJ 08544, USA}

\abstract{We consider the thermal response of a $(3+1)$-dimensional theory with a chiral anomaly on a curved space
motivated by the chiral magnetic effect. 
We find a new phenomenon, called the chiral heat effect, such that the thermal current is induced
transverse to a gradient of the temperature  even on a flat space.
This effect is expected to be observed in QCD experiment as well as the chiral magnetic effect.
We study a similar topological effect on the spacetime with a torsion.
A holographic construction is also discussed with the D3/D7 and the Sakai-Sugimoto models.
}

\preprint{PUPT-2393\\RIKEN-MP-30}

\date{September 2011}

\maketitle

\tableofcontents

\section{Introduction}
Anomalies in quantum field theories have played a crucial role 
since its discovery in the computation of a fermion triangle diagram. 
A classical symmetry is violated quantum mechanically due to an anomaly 
that can be captured by the non-invariance of the path integral measure of
fermions. 
When a theory is classically conformal invariant, there could be the conformal (Weyl)
anomaly in even dimensions which gives the central charges of the theory.
In two-dimensions, the $c$-theorem \cite{Zamolodchikov:1986gt} states that
the central charge monotonically decreases along the renormalization group (RG) flow.
In four-dimensions, there are two central charges, and one of them, named $a$, 
related to the $A$-type anomaly is conjectured to be a monotonically decreasing
function under the RG flow \cite{Cardy:1988cwa}, and it was recently proved by 
Ref.~\cite{Komargodski:2011vj}.
The $c$- (or $a$-) function is an important measure of the number of
degrees of freedom of quantum field theories that is supposed to monotonically
decrease.

A coupling of fermions with the chiral symmetry to a gauge field 
leads to the chiral anomaly, while the fermions gives rise to the mixed
gauge-gravitational anomaly on a curved space. Then, one cannot keep both the chiral symmetry and the gauge invariance (general covariance) 
at the same time.
One can add a counter term to let the chiral anomaly vanish, but it makes the axial current gauge dependent 
or general non-covariant \cite{Fujikawa:2004cx}.

Recent studies of the chiral anomaly reveal that new types of conductivities are induced 
in the presence of background fields \cite{Alekseev:1998ds,Son:2004tq,Metlitski:2005pr,Fukushima:2008xe,Son:2009tf,Landsteiner:2011cp}.
The chiral magnetic effect (CME) is the phenomenon such that an electric current can be parallel to the applied 
magnetic field provided the asymmetry of the chirality between left- and right-handed fermions \cite{Fukushima:2008xe}.
The chiral magnetic conductivity does not receive perturbative corrections because the axial anomaly is 
one-loop exact. Therefore, the CME is not renormalized even at strongly coupled region,
as is implied by the holographic calculations \cite{Yee:2009vw,Rebhan:2009vc,Gorsky:2010xu,Rubakov:2010qi,Gynther:2010ed,Brits:2010pw,Kalaydzhyan:2011vx,Hoyos:2011us}.

Quite similar phenomena are investigated in condensed-matter physics,
\ie~the topological insulator/superconductor \cite{Hasan:2010xy,Hasan:2010hm}.
In particular the time reversal, namely CP symmetric (3+1)-dimensional
topological insulator is based on the $\theta$-term as an effective
action that is also considered in the CME.
In this case, to preserve the time reversal symmetry, the $\theta$-angle
has to be fixed to either $\theta=0$ or $\theta=\pi$, while it is generic for
the CME.
At the boundary of the topological insulator, the domain-wall of the
$\theta$-angle emerges naturally, and thus a massless chiral fermion is
localized there.
This is just a condensed-matter realization of the domain-wall fermion,
which is well studied in lattice gauge theory
\cite{Kaplan:1992bt,Furman:1994ky}.
Such an interesting phenomenon is investigated not only theoretically,
but also experimentally.

In this paper, we consider a $(3+1)$-dimensional fermionic theory with the chiral symmetry on a curved space.\footnote{The mixed gauge-gravitational anomaly is also known to lead to the 
chiral vortical effect in the presence of a magnetic field \cite{Landsteiner:2011cp,Landsteiner:2011iq}.}
We let the axial current conserved, but general non-covariant to introduce the chiral chemical potential.
The calculation of the stress tensor shows that the heat current flows transverse to a gradient of the 
temperature which can be encoded to the off-diagonal components of the background metric.
We call this new phenomenon a ``Chiral Heat Effect'' (CHE) that can happen even on a flat space
with a thermal gradient.
This is a natural counterpart of the CME and a generalization of the surface thermal Hall effect in $(2+1)$ 
dimensions \cite{Ryu:2010ah,Wang:2010xh}.
We also discuss similar topological effect in the presence of torsion and
a possible framework to investigate the CHE in strongly coupling theories in holographic setups.

\section{Chiral heat effect}
Consider a $(3+1)$-dimensional fermionic theory such as QCD on a curved spacetime. 
Suppose the Lagrangian enjoys the chiral symmetry
\begin{align}
	\psi ~\to~ e^{i\alpha \gamma^5/2} \psi \ ,\qquad 
	\bar\psi ~\to~ \bar\psi e^{i\alpha \gamma^5/2}  \ ,
\end{align}
at the classical level, but it is broken at the quantum level due to the chiral anomaly. 
If the theory consists of a massless Dirac fermion, the axial current $j_\m^5$
obeys 
\begin{align}
	\nabla_\m j^{5\m} = -\frac{1}{768\pi^2} \epsilon^{\m\n\rho\sigma}R^\kappa_{~\lambda \m\n} R^\lambda_{~\kappa \rho\sigma} \ .
\end{align}
Now we would like to define a conserved axial current to introduce 
the chiral chemical potential $\m_5$ even on a curved space. 
This is achieved by adding the following functional to
the original action
:\footnote{The Levi-Civita tensor
$\epsilon^{\m\n\rho\sigma}$ is covariant and  is normalized as $\epsilon^{txyz}=-1/\s{-g}$.}
\begin{align}\label{GIaction}
	S =& S_0[\psi,\bar\psi, \mu_5] 
+  \frac{1}{3\cdot 2^8\pi^2} \int d^4 x \s{-g} \, \theta (t,{\vec x})\,\epsilon^{\m\n\rho\sigma}R^\kappa_{~\lambda \m\n} R^\lambda_{~\kappa \rho\sigma} \ ,
\end{align}
where $S_0[\psi,\bar\psi, \mu_5]$ is the action for the fermions with the chiral chemical potential,
and the total action is invariant under the chiral transformation 
with $\theta(t,{\vec x}) \to \theta(t,{\vec x}) + \alpha$. 
The chiral chemical potential term can be removed by the chiral
transformation with $\alpha = \mu_5 t$.
Note that the axial current is no longer covariant after the modification.
This types of the action is considered in the Chern-Simons modification of general 
relativity \cite{Jackiw:2003pm}, and it is topological (a first Pontryagin class) when the function $\theta$ is constant, and the stress-energy tensor 
is zero. It is, however, no longer topological in general and the stress tensor is given by $T^{\m\n} = -\frac{1}{\s{-g}}\frac{\delta S}{\delta g_{\m\n}}$ \cite{Jackiw:2003pm}:
\begin{align}\label{EMtensor}
	T^{\m\n} & = T_0^{\mu\nu}(\mu_5) + T_{H}^{\mu\nu} \ , \nonumber \\
	T_H^{\mu\nu} & = \frac{1}{3\cdot 2^7\pi^2} [ 2 \theta_{;\rho}\,(\epsilon^{\m\rho\sigma\kappa}\, R^{\n}_{~\sigma;\kappa}
	+  \epsilon^{\n\rho\sigma\kappa}\, R^{\m}_{~\sigma;\kappa} ) 
 + \theta_{;\sigma\rho} (\epsilon^{\m\rho\kappa\lambda}R^{\sigma\n}_{~~\kappa\lambda} + \epsilon^{\n\rho\kappa\lambda}R^{\sigma\m}_{~~\kappa\lambda})
	] \ ,
\end{align}
where $T_0^{\mu\nu}(\mu_5)$ is the stress tensor derived from $S_0[\psi,\bar\psi,\mu_5]$ and $``; \kappa''$ stands for the covariant derivative with respect to the index $\kappa$.
In this case, the stress tensor is not conserved
unless the $\theta$ is constant or the Pontryagin density is zero $ \epsilon^{\kappa\lambda\rho\sigma}R^\alpha_{~\beta \kappa\lambda} R^\beta_{~\alpha \rho\sigma}=0$.
In the following discussions, we will consider the situations where the $\theta$ is not constant but the Pontryagin density vanishes,
so general covariance is not broken.
Our discussion is possible even without breaking its general covariance.

When the background is flat space, 
one can show that a thermal
gradient leads to a fluctuation of the metric of the form
\cite{Hartnoll:2009sz,Herzog:2009xv}:
\begin{align}\label{fluctuation}
	i \omega \delta g_{tj} = - \frac{\nabla_j \delta T}{T} \ ,
\end{align}
where the fluctuation of the temperature $\delta T$ is assumed to have a
time dependence $e^{-i\omega t}$.
This means the time dependence is treated in Fourier basis, and thus
(\ref{fluctuation}) shows the temporal derivative of $\delta g_{tj}$
provides a thermal gradient.
This prescription to introduce the thermal gradient was essentially
proposed by Luttinger \cite{PhysRev.135.A1505}.
Note that we introduce this metric deformation just as a perturbation.
Thus we can neglect the back reaction from the matter field.
In other words, we consider that the system is in a thermal equilibrium
state, at least locally.

The thermal current is defined as an operator conjugate the metric, \ie,
the stress-energy tensor unless there are no charged currents, otherwise it is defined by
the difference of the total stress-energy tensor and a charged current:
\begin{align}\label{Thermal_current}
	\langle J_i^T \rangle \equiv T_{ti} - \mu_5 j_i^5 \ .
\end{align}
In our case, the total stress-energy tensor is given by \eqref{EMtensor},
and $T_0^{\mu\nu}(\mu_5)$ includes the current contribution,
which will be canceled by the second term in \eqref{Thermal_current}.
To see this, 
one may introduce the axial background gauge field $A^5_\mu$ with $A^5_t = \mu_5$
coupled to the fermions. The fermion action becomes 
$
S_0[\psi,\bar\psi,\mu_5] \to S_0[\psi,\bar\psi,\mu_5=0] - \int d^4x \sqrt{-g} A^5_\mu j_5^\mu
$
after the gauging the chiral symmetry.
Then, the stress tensor gives rise to the current term
\begin{align}\label{T_0}
	(T_0)_{ti}(\mu_5) = (T_T)_{ti} +  \mu_5 j_i^5 \ ,
\end{align}
where $T_T^{\mu\nu} =  T_0^{\mu\nu}(\mu_5 = 0) + \frac{\mu_5 j_5^0}{2} g^{\mu\nu}$ .
Substituting \eqref{EMtensor} and \eqref{T_0} into the definition of the heat current \eqref{Thermal_current}, one obtains 
\begin{align}\label{HC}
	\langle J_i^T \rangle = T^T_{ti} + T^H_{ti} \ .
\end{align}
The first term is the heat current from the fermionic action $S_0[\psi,\bar\psi,\mu_5]$ which does not have Hall conductivities.
The second term coming from the additional action in \eqref{GIaction} can have Hall conductivities as we will see below.

Let us consider an example. Suppose the space coordinates are flat and the temperature depend on $x$ direction. Using the diffeomorphism, the temperature can be set to constant, while there appears the off-diagonal component $\delta g_{tx}(x)$ in the spatial metric from the relation \eqref{fluctuation}. 
It gives rise to the heat current along $x$ direction $\langle J_x^T \rangle = T_{tx}^T$ 
and we obtain the usual heat conductivity: $\kappa_{xx} = T_{tx}^T/(\partial_x \delta T)$.
Now our interest is focused on the Hall conductivity. 
The stress tensor $T^H_{\mu\nu}$ \eqref{EMtensor} can lead the heat current
transverse to the $x$ direction if $\theta$ depends on $z$ direction
\begin{align}\label{CHE}
	\langle J^T_y \rangle = \frac{-i\omega}{3\cdot 2^7\pi^2}
	\partial_z \theta\, \partial_x^2 \delta g_{tx}(x)  \ ,
\end{align}
where we suppress $e^{-i\omega t}$ for simplicity. 
Remark this effect shows a non-linear transport phenomenon, including
higher derivative terms.
Thus it cannot be characterized by the usual linear transport coefficient,
\ie, the thermal conductivity.
Similar situation occurs in the theory of the topological insulator
\cite{Qi:2008ew}.

Other interesting situation happens in the presence of a time-dependent $\theta$.  
If a gradient of the temperature depends on $x$ and $y$ coordinates, 
one can convert it into the fluctuation of the metric components $\delta g_{tx}$ and $\delta g_{ty}$.
Then a heat current is induced by the thermal distribution in the perpendicular plane:
\begin{align}\label{CHEcp}
	\langle J^T_z \rangle & = -\frac{1}{3\cdot 2^8\pi^2} \partial_t \theta\, [ \partial_y^3 \delta g_{tx}(x,y) 
	-\partial_x\partial_y^2 \delta g_{ty}(x,y) 
 + \partial_x^2\partial_y \delta g_{tx}(x,y) 
	-\partial_x^3 \delta g_{ty}(x,y) ] \ .
\end{align}
The time-dependence of $\theta$ is related as $\theta = \mu_5 t$ with the chiral chemical potential $\m_5$
as discussed in the context of 
the chiral magnetic effect \cite{Fukushima:2008xe}.
The mechanism of this thermal Hall effect is analogous to the chiral magnetic effect.
The role of the magnetic field of the CME is played by the fluctuation of the temperature which 
turns out to be a curvature in the spatial directions by using the diffeomorphism. Both of them are triggered by the 
chiral anomaly in the presence of the chiral chemical potential.
We would like to call the above phenomenon the ``Chiral Heat Effect''
(CHE) with emphasis on the similarity to the CME.
The CHE can happen even on a flat space with a thermal gradient in the presence of 
the time-dependent theta angle or the chiral chemical potential.
This is also a non-linear effect with respect to the fluctuation of the
background temperature.
Note that the right hand side of (\ref{CHEcp}) has a singular behavior
$i/\omega$ in the zero-frequency limit $\omega \to 0$.
This singularity implies the conductivity includes the $\delta$-function
term, \eg~$\sigma(\omega) = \pi D \delta(\omega) + \sigma_{\rm reg}(\omega)$.
The coefficient $D$ is called the Drude weight, and this kind of
singularity is generally observed in the ballistic system.
This behavior is regarded as a result of the translation symmetry of the system.


When we assume the time reversal symmetry of the system, the $\theta$-angle
 is fixed to either $\theta = 0$ or $\theta = \pi$, because it is
inverted as $\theta \to - \theta$ (mod $2\pi$) under the time reversal operation.
The state with $\theta = \pi$ is topologically non-trivial, 
while it is trivial for $\theta=0$.
This means the $\theta$-angle has to jump from $\theta = \pi$ to $\theta =
0$ at the boundary of the topological state.
Therefore we can obtain the gravitational Chern-Simons term on a
domain-wall between topologically non-trivial and trivial states,
\begin{equation}
 S
  = 
  \frac{1}{384 \pi} \int d^3 x \,\det e \, \epsilon^{\m\n\rho}\, {\rm Tr}
  \left(
   \omega_\mu \partial_\n \omega_\rho + \frac{2}{3} \omega_\m \omega_\n \omega_\rho
  \right) ,
\end{equation}
where $e^a_\m$ and $\omega^{~a}_{\m ~b}$ are triad and spin connection, respectively, and 
the trace is taken for the frame indices $a$.\footnote{Here
the $\epsilon^{\m\n\rho}$ is an anti-symmetric tensor with respect to 
all indices, and is normalized to $\epsilon^{txy}=-1/\det e$.}
This effective action yields the transverse thermal response at the
boundary as shown in (\ref{CHE}).
It is discussed that this action leads to a half-integer quantization of the
thermal Hall conductivity \cite{Ryu:2010ah,Wang:2010xh}.
A similar anomalous quantization is observed for the Hall current and
spin Hall current at the boundary of the topological insulators
\cite{PhysRevB.50.7526,Qi:2008ew,PhysRevLett.102.196804}.

\section{Topological effect with torsion}
Let us comment on another possibility of theoretical generalization:
we now investigate a similar topological effect in the presence of torsion.
We consider the Nieh-Yan topological
action \cite{Nieh:1981ww}, which is defined as
\begin{eqnarray}
 S & = & \frac{1}{32\pi^2} \frac{2}{\ell^2} \int d^4 x \det e\, \theta(t,{\vec x})\, 
 \epsilon^{\m\n\rho\sigma}
  \left(
   T_{~\m\n}^a T_{a \rho\sigma}
   - R_{ab\m\n} e^a_{~\rho} e^b_{~\sigma}
  \right) \ ,
  \label{NYaction}
\end{eqnarray}
where the torsion and Riemann tensors are given by $T^a = de^a + \omega^{a}_{~b}
 \wedge e^b$ and $R^a_{~b} = d\omega^a_{~b} + \omega^a_{~c}\wedge \omega^c_{~b}$,
 respectively. 
Although the same character stands for the stress and torsion
tensors, one can identify it by the number of suffixes.
Note that the dimensionful constant $\ell$ is required for the action
 (\ref{NYaction}), which leads to controversy on the topological origin
 of this action (see, for example, Refs.~\cite{Chandia:1997hu,Kreimer:1999yp,Chandia:1999az,Li:1999ue,Mielke:2006gi,Mielke:2009zz}).

Here we consider the stress tensor given by $\frac{1}{\det
e}\frac{\delta S}{\delta e^a_{~\m}}$, instead of the standard one
$\frac{-2}{\sqrt{-g}}\frac{\delta S}{\delta g_{\mu\nu}}$, for the
Nieh-Yan action,
\begin{equation}
 \left\langle T_a^{~\m} \right\rangle 
  = 
  \frac{1}{8\pi^2 \ell^2}
  \epsilon^{\m\n\rho\sigma} \, \partial_\n \theta 
  \left( 
   \partial_\rho e_{a \sigma} + \omega_{ab \rho} e^b_{~\sigma}
  \right) \ .
  \label{VEtensor}
\end{equation}
We can freely switch between the frame and spacetime coordinates within the linear elasticity theory.
Introducing a displacement field $u^a(x)$, 
the stress tensor can be written with the strain tensor
$u_{\mu\nu} = (\partial_\mu u_\nu + \partial_\nu u_\mu)/2$,
\begin{equation}
 \left\langle T^{\m\n} \right\rangle = \Lambda^{\m\n\rho\sigma} u_{\rho\sigma} 
  + \eta^{\m\n\rho\sigma} \dot{u}_{\rho\sigma}\ ,
  \label{VEtensor2}
\end{equation}
where $\Lambda$, $\eta$ are the elasticity and viscosity tensors.
The anti-symmetric part involves a dissipationless viscoelastic
response, $\eta_{\rm H}^{\m\n\rho\sigma} = - \eta_{\rm
H}^{\rho\sigma\m\n}$, which is called the Hall viscosity
\cite{Avron:1995fg,Read:2008rn,Kimura:2010yi,Hughes:2011hv,Nicolis:2011ey,Saremi:2011ab,Hoyos:2011ez},
while the symmetric part corresponds to the dissipative viscosity.
The triad can be written as
$e^a_{~\m} = \delta^a_{~\m} + \partial_\m u^a$, where $\partial_\mu u^a$
is a distortion tensor.
Thus we can observe the dissipationless viscosity from
(\ref{VEtensor}), when the $\theta$-angle is dependent on $z$,
\begin{equation}
 \left\langle T_a^{~i} \right\rangle =
  \frac{1}{8\pi^2 \ell^2}
  \epsilon^{ij} \, \partial_z \theta \, \dot{u}_{a j}\ .
\end{equation}
Here we omit the contribution from the spin connection.
When we consider the domain-wall between $\theta=0$ and $\theta=\pi$,
namely the boundary of the topological insulator, the Hall viscosity is
given by
\begin{equation}
 \eta_{\rm H} = \frac{\hbar}{8\pi \ell^2}\ .
\end{equation}

We then study another configuration, which is analogous to the CME: the $t$
dependence of the $\theta$-angle yields
\begin{equation}
 \left\langle T_a^{~i} \right\rangle = \frac{1}{8\pi^2 \ell^2}
  \epsilon^{ijk} \, \partial_t \theta \, \partial_j e_{a k}\ .
  \label{CTE}
\end{equation}
This shows the momentum current is proportional to the Burgers vector in
the transverse plane, ${\bf b}^a = \epsilon^{ij} \partial_i
e^a_{~j}$, which can be interpreted as a flux of torsion.

Let us remark a relation to the CME and the chiral vortical effect (CVE).
All these effects are based on the CP-odd topological terms, and thus
coming from rotations of vector fields: the vector potential for the CME,
the carrier current for the CVE and the triad for the torsional effect (\ref{CTE}).
On the other hand, the standard response is linearly dependent on the
triad itself and its temporal derivative, as shown in the first and
second terms in (\ref{VEtensor2}), respectively, since it can
be identified with the distortion tensor up to higher order corrections.

\section{Discussions}


In this paper, we considered the chiral heat effect such that the heat current flows perpendicular to the direction of the thermal gradient in the presence of the chiral (or mixed gauge-gravitational) anomaly in a field theory. 
Since anomaly is supposed to be independent of the gauge coupling, we may well be able to observe this effect even in strongly coupled theories.
Before concluding this paper, we would like to suggest a possible framework in this direction by using the AdS/CFT correspondence.

Consider $N_f$ probe D7-branes on the AdS$_5$-Schwarzschild black hole times $S^5$ background.\footnote{We
use the convention of Ref.~\cite{Myers:1999ps} where the Newton constant is chosen 
such that the dilaton vanishes asymptotically. The RR-charge density, then, equals to the D-brane
tension: $\mu_p = T_p = \frac{1}{(2\pi)^p \alpha'^{\frac{p+1}{2}}g_s}$\ .}
The dual field theory is the $\CN=2$ super Yang-Mills theory with $N_f$ fundamental hypermultiplets.
The Wess-Zumino term gives rise to the following on  the background with an RR four-form \cite{Green:1996dd,Cheung:1997az}
\begin{align}
	S_{WZ} = \frac{\mu_7 N_f (4\pi^2 \alpha')^2}{2\cdot 24\cdot 8\pi^2} 
	\int_{\rm D7} C_{4} \we {\rm Tr}(R_T\we R_T - R_N\we R_N) \ .
\end{align}
The $R_T$ and $R_N$ denote the Riemann tensors of the tangent and the normal frame
of the D-branes, respectively.
We will use the background metric of the form
\begin{align}\label{metric}
	ds^2 & = \frac{\gamma^2 \rho^2}{2}\left(-\frac{f^2(\rho)}{H(\rho)}dt^2 + H(\rho) d{\vec x}^2\right) 
 + \frac{1}{\rho^2} (dr^2 + r^2 ds^2_{S^3} + dR^2 + R^2 d\phi^2) \ ,
\end{align}
where we have set the AdS radius to $L=1$. Then we can always convert the 
string length with the 't Hooft coupling: $\alpha'^{-2} = \lambda = 4\pi g_s N_c$ where
$N_c$ is the rank of the gauge group $U(N_c)$ of the dual field theory.
The RR four-form $C_{4}$ is given by
\begin{align}
	C_{4}= \frac{\rho^4\gamma^4 H^2(\rho)}{4} {\rm vol}_{{\mathbb R}^{3,1}} - \frac{r^4}{\rho^4}d\phi
	\we {\rm vol}_{S^3} \ .
\end{align}

Let the D7-branes spread over the AdS$_5$ spacetime and wrap on the three-sphere inside
the $S^5$. 
Then we obtain the Wess-Zumino term of the form
\begin{align}\label{WZ}
	S_{WZ} = - \frac{N_c N_f}{3\cdot 2^7\cdot \pi^2} \int_{AdS_5} \frac{r^4}{\rho^4} d\phi \we
	 {\rm Tr}(R_T\we R_T - R_N\we R_N) \ .
\end{align}
Note that this term comes from higher derivative terms in the WZ action, but it is the order $\CO(N_c N_f)$ quantity.
If the $\phi$ depends on the space coordinates,
the first term will describe the chiral heat effect holographically
along the lines of Ref.~\cite{Hoyos:2011us} where the chiral magnetic effect was 
investigated replacing the curvature with the worldvolume gauge field in the action \eqref{WZ}.
The $\phi$ will be identified with $\theta$ in Eq.\,\eqref{GIaction} since it acts on hypermultiplets in the dual $\CN=2$ 
SYM theory as a $U(1)_R$ symmetry: $(q,\bar q) \to (e^{i\phi\gamma^5/2}q, \bar q e^{i\phi\gamma^5/2})$.
Then the holographic computation would reproduce the CHE
\eqref{CHEcp} for $N_f$ fermions in fundamental representation of $U(N_c)$. 
The second term including the $R_N$, which is irrelevant to our present discussion, is related with $U(1)_R SU(2)_R^2$ and $U(1)_R SU(2)_L^2$ anomalies
\cite{Aharony:2007dj} where the $SU(2)_R$ and the $SU(2)_L$ act on the dual $\CN =2$ theory 
as the R-symmetry and the global symmetry rotating the adjoint scalars, respectively.

\bigskip
Another way to construct the CHE holographically would be the use of the D4/D8 system
that contains D8-branes wrapped on $S^4$ like the Sakai-Sugimoto model\cite{Sakai:2004cn,Sakai:2005yt}.
The essential logic is the same as before, and we consider the WZ term of D8-branes
in the presence of a RR one-form $C_1$ by putting D0-branes:
\begin{align}\label{D8WZ}
	S_{WZ} = \int_{\rm D8} C_1 \we F \we F \we {\rm Tr}(R_T\we R_T - R_N\we R_N) \ .
\end{align}
Given a non-zero instanton number $\int_{S^4} F\wedge F \neq 0$ on D8-branes which represents the number of D4-branes
dissolved on the $S^4$ inside the D8, one obtains a similar effective action to \eqref{WZ}.
Here the RR one-form plays a role of the derivative of the $\theta$ in the dual field theory.


\bigskip
Although it seems hard to construct a thermally fluctuating background around 
the solution \eqref{metric},
it would be interesting to compute the thermal current in the strongly-coupled 
theories dual to these holographic models and 
check that the current does not depend on the gauge coupling 
due to the non-renormalizability of the chiral anomaly.
The future work should be devoted to a concrete calculation of the
current with these holographic systems.

 \vspace{.5cm}
 \centerline{\bf Acknowledgements}
We are grateful to M.\,Kaminski, C.\,Hoyos, A.\,O'Bannon, Y.\,Hidaka,
A.\,Shitade, A.\,Tanaka and H.\,Verlinde for valuable discussions.
T.\,K. is supported by Grant-in-Aid for JSPS Fellows.
The work of T.\,N. was supported in part by the US NSF under Grants No.\,PHY-0844827 and
PHY-0756966.

\bibliographystyle{ytphys}
\bibliography{THE}

\end{document}